\documentstyle[12pt,epsf]{article}

\newcommand{\beq}{\begin{equation}}
\newcommand{\eeq}{\end{equation}}
\newcommand{\beqa}{\begin{eqnarray}}
\newcommand{\eeqa}{\end{eqnarray}}
\newcommand{\boldtau}{\mbox{\boldmath $\tau$}}

\newcommand{\boldpi}{\mbox{\boldmath $\pi$}}
\newcommand{\boldsig}{\mbox{\boldmath $\sigma$}}
\newcommand{\boldr}{\mbox{\boldmath $r$}}

\begin{document}

\begin{titlepage}


\hfill{KRL MAP-267}


\hfill{NT@UW-00-15}

\vspace{1.0cm}

\begin{center}
{\Large {\bf Charge Symmetry  Violation in $pn\rightarrow d \pi^0$ \\
    as a Test of Chiral Effective Field Theory}}

\vspace{1.2cm}

{\large 
U. van Kolck $^{a,b,}$\footnote{{\tt vankolck@krl.caltech.edu}},
J.A. Niskanen $^{c,}$\footnote{{\tt jouni.niskanen@helsinki.fi}},
and
G.A. Miller $^{b,}$\footnote{{\tt miller@phys.washington.edu}}}

\vspace{1.2cm}
{\it
$^a$ Kellogg Radiation Laboratory, 106-38,
 California Institute of Technology, \\
 Pasadena, CA 91125, USA
~\\$^b$ Department of Physics,
 University of Washington, \\
 Seattle, WA 98195, USA
~\\$^c$ Department of Physics,
 University of Helsinki, \\
 FIN-00014 Helsinki, Finland}
\end{center}

\vspace{1cm}

\begin{abstract}
Chiral effective field theory predicts a specific charge symmetry violating
amplitude for
pion production. This term is shown to provide
the  dominant 
contribution to the forward-backward 
asymmetry in the
 angular distribution for  
the reaction $pn\to d\pi^0$, for reasonable
 values of the mass difference between down  and up quarks, 
$\delta m_N$.  Using a value $\delta m_N\approx$ 3 MeV leads
to a prediction of a 10 standard deviation effect for a TRIUMF
 experiment.

\end{abstract}

\vspace{2cm}
\vfill
\end{titlepage}
Charge symmetry is an approximate symmetry of QCD. 
If one ignores electromagnetic
corrections and the mass difference between $u$ and $d$ quarks, then the
Lagrangian is invariant under the interchange of $u\leftrightarrow d$.
Isospin invariance implies charge symmetry, but the converse is not true.

Much has been learned about charge symmetry breaking (CSB); see the
reviews \cite{reviews,jerry}. 
The general importance of the quark mass difference
effect was stressed in Ref. \cite{jerry}. 
Our purpose here is to  
use chiral effective field theory to make predictions, based on QCD,
for the reaction $np\to d\pi^0$.

The use of chiral Lagrangians, which employ  
hadronic degrees of freedom,
represents a serious effort to provide a rigorous and model-independent
methodology to use QCD to
make predictions at relatively low energies \cite{book1,book2,vk}.
This is because the most general Lagrangian 
which respects unitarity, has correct properties
under cluster decomposition and the same symmetries as QCD, should give
results that encompass the results from QCD proper \cite{W79}.
Predictive power is retained because at low energies
an expansion in momentum can be formulated using power counting
arguments.

We wish to exploit the feature that  chiral symmetry makes predictions for 
effects 
that stem from the small but {\it explicit} breaking of chiral symmetry
generated by the quark masses.
Consider the proton-neutron mass difference  $m_n-m_p$,
which includes a contribution $\delta m_{N}$ 
from the quark mass difference $m_d-m_u \equiv \varepsilon (m_d+m_u)$
($\varepsilon \sim 1/3$),
and another contribution $\bar\delta m_{N}$ 
from electromagnetic effects. Thus  $m_n-m_p=\delta m_{N}+\bar\delta m_{N}$.
Neglecting strangeness, chiral symmetry is essentially an $SO(4)$ internal
symmetry. 
One can show that the quark-mass-difference term in the QCD Lagrangian
behaves under $SO(4)$ as the third component of an $SO(4)$ vector
\cite{ivwei1,me,weinagain}.
Therefore, in the effective hadronic theory, all isospin-violating interactions
generated by  the quark mass difference  break
$SO(4)$ as third components of (tensor products of) $SO(4)$ vectors.
The operators of interest here involve the nucleon $N$ and pion 
$\boldpi$ fields. 
The leading isospin-violating term at low energies coming 
from the quark mass difference is the
third component of an $SO(4)$ vector \cite{me,weinagain}:
\begin{equation}
 {\cal L}_{\rm qm}^{(1)}= 
 \frac{\delta m_{N}}{2}
                       \left(N^\dagger \tau_{3}N
                        -\frac{2}{DF_{\pi}^{2}}
                         N^\dagger \pi_{3} \boldpi \cdot \boldtau N \right),  
                                                               \label{iv24}
\end{equation}
\noindent
where $\boldtau$ represents the Pauli matrices in isospin space,
$F_\pi=186$ MeV is the pion decay constant, and
$D=1+\boldpi^2/F_\pi^2$. We use $D=1$ here.
Eq. (\ref{iv24}) represents the isospin-violating part of the 
nucleon sigma term. 

Furthermore, one can show that quark interactions
generated by (``hard'') photon exchange
break $SO(4)$ as the 34 component of an $SO(4)$ antisymmetric rank-2 tensor
\cite{me}. Thus the low-energy effective theory must include
 isospin-violating interactions that break
$SO(4)$ as 34 components of (tensor products of) $SO(4)$
antisymmetric rank-2 tensors.
The leading isospin-violating term of this kind at low energies was
shown to be \cite{me}:
\begin{equation}
 {\cal L}_{\rm hp}^{(-1)} =  
 \frac{\bar{\delta} m_{N}}{2}
                       \left(N^\dagger \tau_{3}N
                        +\frac{2}{DF_{\pi}^{2}}
      N^\dagger (\pi_{3} \boldpi\cdot \boldtau -\boldpi^2 \tau_3) N\right).
                                                               \label{iv25}
\end{equation}
\noindent

Observe that
chiral symmetry links
the first terms in Eqs. (\ref{iv24},\ref{iv25}), which
are
contributions to the nucleon mass difference,
to the second terms, which are isospin-violating 
pion-nucleon interactions. 
Since the parameter $F_\pi$ is determined from another process (pion decay), 
these pion-nucleon interactions
{\it must  exist} with the given strength
if our understanding of QCD is not totally flawed.

Let us discuss the values of the  parameters, $\delta m_{N}$ and $\bar\delta
m_{N}$, which are not
determined by chiral symmetry. 
The sum of these terms is fixed by the nucleon mass difference
$m_n-m_p =1.3$ MeV. 
Dimensional analysis suggests their separate order of magnitude
in terms of the natural QCD mass scale, $M_{QCD}\sim 1$ GeV.
We know $\delta m_{N}\propto 
m_d-m_u$ and $m_\pi^2\propto  m_d+m_u$,
so
$\delta m_{N}= O(\varepsilon m_\pi^2/M_{QCD})\sim  7$ MeV.
The electromagnetic term
$\bar\delta m_{N}$, long known to be negative \cite{zee},
should be proportional to the fine structure
constant $\alpha$, so that
$\bar\delta m _N =- O(\alpha M_{QCD}/\pi)\sim -2$ MeV.
Estimates such as these cannot be trusted to  better
than a factor of a few. 
More precise estimates can only be made at present 
with model-dependent assumptions.
Most models estimate that
$\bar\delta m_{N}\approx -\alpha/R_N \simeq -1.5$ MeV
\cite{jerry},
in which case  $\delta m_{N} \simeq 3$ MeV.
We will refer to these values, which are also consistent with the results of
Gasser and Leutwyler \cite{gl}, as ``quark model estimates'' .

Verifying the existence of the terms of Eqs.~(\ref{iv24},\ref{iv25}) is
of high interest, as is determining the parameters. 
The simplest procedure would be to study
pion-nucleon scattering close to threshold \cite{ivwei1,me,weinagain,nadia}.
The lack of $\pi^0$ beams requires the comparison of charge
exchange and  elastic scattering processes.
If one neglects electromagnetic effects and
 the kinematic 
 dependence on the threshold values of the energy   on the masses, 
the leading-order chiral Lagrangian gives for the 
triangle discrepancy among the pion-nucleon amplitudes $T$:
 \begin{equation}
D= -2 \frac{T(\pi^{+}p\rightarrow\pi^{+}p) -T(\pi^{-}p\rightarrow\pi^{-}p)
   -\sqrt{2} T(\pi^{-}p\rightarrow\pi^{0}n)}
        { T(\pi^{+}p\rightarrow\pi^{+}p) -T(\pi^{-}p\rightarrow\pi^{-}p)
   +\sqrt{2} T(\pi^{-}p\rightarrow\pi^{0}n)}
   \approx \;-\;\frac{\delta m_N-\bar{\delta} m_{N}}{2m_\pi}.
\label{deq}\end{equation}
The quantity $D$ is found \cite{gibbs} to be about 6\%, 
leading to a very large {\bf negative} 
value of $\delta m_N-\bar{\delta} m_{N}\approx -17 $ MeV. 
If the 6\% result is correct, there seems to be a serious problem
with our current understanding of low-energy QCD. 
However, electromagnetic effects are very important at low energies as are
the kinematic difference in threshold energies. 
Kinematic effects and sub-leading strong interactions have been studied
in Ref. \cite{nadia}, but 
a precise
accounting for the electromagnetic interactions presents a
very serious challenge to theorists. 
A small shift in
the normalization could be reflected in a huge change in $D$. It seems that 
a  successful determination of 
$\delta m_N$ and $\bar{\delta} m_{N}$ from $\pi N$ scattering 
will 
require a strong theoretical and experimental effort.

The biggest problem in all of this is the lack of a pion beam.
The reaction $np\to d\pi^0$ is relevant because  it involves the 
production of a  virtual pion from
one  nucleon (``pion bremsstrahlung''), followed by
rescattering on the second. 
Charge symmetry predicts
that the angular distribution is symmetric about 90$^\circ$ in the
center-of-mass.  Any asymmetry must be caused by
charge-symmetry-violating effects. 
We here examine the 
integrated forward-backward asymmetry in the center of mass
of the reaction $np \rightarrow d \pi^0$,
\begin{equation}
A_{fb}=\frac{\int_0^{\frac{\pi}{2}} 
           d\Omega \; [\sigma(\theta)-\sigma(\pi-\theta)]}
{\int_0^\pi d\Omega \; \sigma(\theta)}.
              \label{asym}
\end{equation}
Our interest in this particular observable is motivated by
an experiment in progress at TRIUMF \cite{allena}
which aims  to measure this asymmetry 
close to threshold, $E_{lab}=279.5$ MeV or $\eta= q_\pi/m_\pi=0.17$,
with a precision of $\pm$0.06\%
\footnote{The definition of the asymmetry $A_{fb}$
used here is $1/2$ of the definition used in Ref. \cite{allena}.}.

$A_{fb}$ arises from  an interference between $s$- and 
$p$-wave pions and thus decreases as one approaches threshold energies. 
Nevertheless, there are advantages to considering this pion production reaction
at threshold kinematics.
Near threshold
one minimizes the four-momentum transferred $k$, enhancing 
isospin violation relative to chiral invariance, because
chiral (thus isospin) conserving contributions to $\pi N$ rescattering
vanish with the pions' four-momenta.
Right at threshold the leading $\pi N$ rescattering mechanism
(Weinberg-Tomozawa term \cite{weitom}) goes as the sum of the
pions' energies, $\omega_q +\omega_k$;
the real pion  has as small energy
as possible, $\omega_q \simeq m_\pi$, 
while the typical energy of the virtual
pion is just $\omega_k\sim m_\pi/2$. 
In pion production the three-momentum $p\approx
\sqrt{m_Nm_\pi}$
of the virtual pion is relatively large, which slows the convergence of 
a momentum expansion. However, 
there is a trade-off here, since
the large momentum reduces the importance of electromagnetic effects.
Note, also, that 
at any given order in a momentum expansion in $p/M_{QCD}$,
the number of relevant isospin-breaking contributions is more limited
than isospin-conserving ones. This  leads  to an isolation
of $\delta m_N$ and $\bar\delta m_N$ effects. Even if other mechanisms
({\it e.g.} $\pi$-$\eta$ mixing) are also relevant, they are
not expected to exactly cancel among themselves. Thus, a model adjusted
to correctly describe the total cross-section might provide a 
reliable estimate 
of the overall size of the $\delta m_N$ and $\bar\delta m_N$
contributions to the asymmetry; and these estimates would 
 be confronted with data.

The most recent  calculation of the near-threshold
asymmetry \cite{jouni}  includes effects of the 
nucleon-mass splitting on the $\pi N$ vertex      function,
which influences both the pion production amplitude and the 
nucleon-nucleon potential, 
and the far larger influence
of $\pi$-$\eta$-$\eta '$ mixing,
both in  the pion-production kernel and
in initial and final interactions between nucleons. This calculation 
 does not include the seagull interactions (\ref{iv24},\ref{iv25})
 proportional 
to $\delta m_N$ and $\bar\delta m_N$.
The isospin-conserving interactions are included in a
nucleon-delta coupled-channel model that  describes
both pion production and
$NN$ phase shifts to an accuracy of a few degrees from
threshold over the delta region \cite{morejouni}.
Close to threshold $\pi$-$\eta$ mixing
is  the dominant effect, 
taking 
place both in pion $s$ and  $p$  waves.

\begin{figure}[t]
\epsfxsize=15cm
\centerline{\epsffile{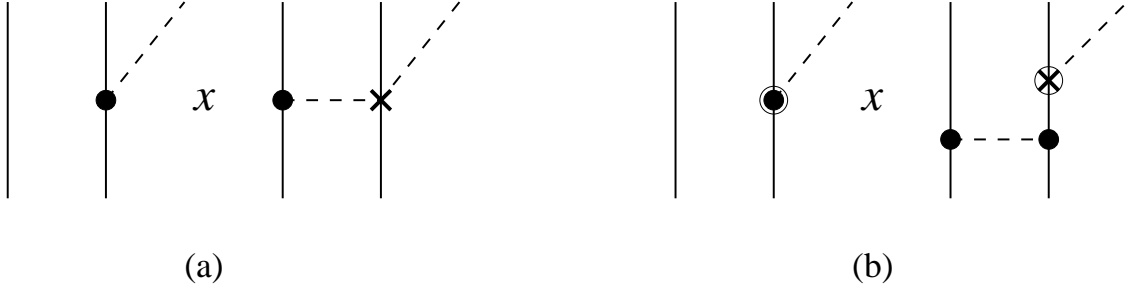}}
\caption{Contributions to $A_{fb}$: 
(a) ${\cal M}_{\rm qm}$; (b) ${\cal M}_{\pi\eta}$.
Here nucleon (pion) propagation is represented by 
a solid (dashed) line;
and pion emission/absorption is denoted by
a (circled) solid dot in case of an isospin-conserving $p$ ($s$) wave,
and a (circled) cross in case of an isospin-violating $s$ ($p$) wave.}
\label{fig1}
\end{figure}

Here we estimate, using both
a power counting argument and a numerical evaluation,
the effect of $\delta m_N$ and $\bar\delta m_N$
in the asymmetry (\ref{asym}) close to threshold.
These mechanisms appear in the $s$ wave,
and can be relatively large because the isospin-conserving
$p$ wave 
amplitude is not vanishingly  small when compared
to the isospin-conserving $s$ wave, even below
300 MeV.
A  power counting argument suggests immediately that the
contributions from Eqs.~(\ref{iv24},\ref{iv25}) should be dominant.
Let us take the contribution from the $\delta m_N$ seagull, Fig. \ref{fig1}a,
as an example.
The isospin-conserving $p$ wave emission is proportional
to $q/F_\pi$, while the isospin-violating $s$ wave
gets a
$p/F_\pi$ from the virtual pion emission,
$1/p^2$ from the pion propagator, and 
$\delta m_N/F_\pi^2$
from the seagull vertex.
The product of the two amplitudes 
is then expected to be
\begin{equation}
{\cal M}_{\rm qm}\sim
\varepsilon \left(\frac{m_\pi}{m_N}\right)^{3/2} \frac{q}{F_\pi^4}, 
\label{qmd}\end{equation}
in which the notation qm is used to denote the effects of
Eq. (\ref{iv24}).
On the other hand, 
the contribution from $\pi$-$\eta$ mixing is smaller by a factor
of $m_\pi/m_N$ near threshold.
This can be seen by integrating out the $\eta$. This produces
an isospin-violating pion-nucleon coupling of size
$\beta q/F_\pi$ in the $p$ wave and 
$\beta \omega_q p/F_\pi m_N$ in the $s$ wave,
where  $\beta \sim \varepsilon m_\pi^2/m_N^2$
\cite{friargoldman}.
Let us consider now isospin violation in the $p$-wave pion emission, 
Fig. \ref{fig1}b.
The isospin-conserving $s$-wave emission is proportional
to $\omega_q p/F_\pi m_N$,
while the isospin-violating $p$ wave
gets a
$p/F_\pi$ from the virtual pion emission,
$1/p^2$ from the pion propagator, 
another $p/F_\pi$ from the virtual pion absorption, 
$1/\omega_q$ from the nucleon propagator, 
and $\beta q/F_\pi$  from the real pion emission.
The product of the two amplitudes is then expected to be
\beq {\cal M}_{\pi\eta}\sim
\varepsilon \left(\frac{m_\pi}{m_N}\right)^{5/2} \frac{q}{F_\pi^4}. 
\label{pieta}\eeq
The same estimate holds for the interference
between isospin-conserving $p$ wave and isospin-violating $s$ wave. 
A comparison of Eqs. (\ref{qmd}) and  (\ref{pieta}) reveals the ratio
$m_\pi/m_N$ between the two effects.
Note that the counting rules used here  are consistent with those of
previous works in pion production\cite{counting}.

We calculate $A_{fb}$ following the procedure
described in Ref. \cite{jouni} and  earlier references.
Before presenting the results, we display aspects of the relevant operators.
Denoting the two nucleons with superscripts $^{(1)}$ and $^{(2)}$,
the spin-isospin structure of the isospin-violating rescattering is (at
threshold,
and neglecting terms of order $m_\pi/m_N$ )
\begin{eqnarray}
{\cal O}_{\rm qm} &=& -\frac{f}{4\pi\mu F_\pi^2 \sqrt{2\mu}}
[\delta m_N (\boldtau^{(1)}\cdot \boldtau^{(2)} 
+\tau_3^{(1)}\tau_3^{(2)})  
- \bar\delta m_N (\boldtau^{(1)}\cdot \boldtau^{(2)} 
-\tau_3^{(1)}\tau_3^{(2)})]  \nonumber  \\
 & & (\boldsig^{(1)} - \boldsig^{(2)})
\cdot (-i\hat{\boldr} ) \frac{d}{dr} \left(\frac{e^{-\mu' r}}{r} \right),
\label{odeltam}
\end{eqnarray}
where 
$\mu$ is the charged pion mass, 
$\mu'=\sqrt{3}\mu/2$,
and $f$ is the pion-nucleon coupling constant
($f=g_A \mu/F_\pi$ in leading order;
$f^2/4\pi = 0.075$ is used here). The isospin  operator allows transitions
from states with isospin $I=0$ to the deuteron, also $I=0$:
\beq
\langle I=0|\boldtau^{(1)}\cdot \boldtau^{(2)} \pm \tau_3^{(1)}\tau_3^{(2)}|
I=0\rangle
=-3\;\mp\;1,
\eeq
where only the isospin is indicated. Thus there is a large matrix element.
If we do the spin-angle evaluation of the matrix element of 
${\cal O}_{\rm qm}$ between
$^1P_1$ and $^3S_1$ (for the deuteron) states we find
\beq
\langle^3S_1 |{\cal O}_{\rm qm}(r)| ^1P_1\rangle=
 \frac{2f}{\pi\mu F_\pi^2 \sqrt{3}
   \sqrt{2\mu}} ( \delta m_N - \frac{\bar\delta m_N}{2})(-i)
 \frac{d}{dr} \left(\frac{e^{-\mu' r}}{r}\right)
 .\eeq

We emphasize that
the asymmetry in pion production thus depends on a different
combination of   $\delta m_N$ and $\bar\delta m_N$
than either the total nucleon mass splitting or
the isospin violation related to charge-exchange pion-nucleon scattering.
Under the assumption that we have the
isospin-conserving mechanisms of pion production under control,
$A_{fb}$ can be used in conjunction with $m_n-m_p$ to extract
 $\delta m_N$ and $\bar\delta m_N$.

The above operator may be compared with the operator
involving phenomenological isospin-conserving, on-shell pion  rescattering,
which in leading order is given by the Weinberg-Tomozawa
term \cite{weitom}, sometimes written in terms of 
$\lambda_2=\mu^2/4\pi F_\pi^2 \approx0.045$.
If we consider threshold kinematics,  and only the matrix element between
$^3P_1$ and $^3S_1$ states, we obtain
\cite{koltun}
\begin{equation}
{\cal O}_{\rm ic} =  -\frac{3 f} {2 \mu^2  \sqrt{2\mu}}
\lambda_{2}
 \;i(\boldtau^{(1)}\times\boldtau^{(2)})_3
(\boldsig^{(1)} +\boldsig^{(2)} )
\cdot (-i\hat{\boldr} ) \frac{d}{dr} \left(\frac{e^{-\mu' r}}{r} \right),
\end{equation}
so that
\beq
\langle^3S_1 |{\cal O}_{\rm ic}(r)| ^3P_1\rangle=
6\sqrt{2\over3} \frac{  f} {\mu^2  \sqrt{2\mu}}
\lambda_2
(-i) \frac{d}{dr} \left(\frac{e^{-\mu' r}}{r} \right)
.\eeq
Assuming 
 that the initial  $np$ $^3P_1$ and $^1P_1$ wave functions are the
same
(the phase shifts differ by only $\sim 10\%$ at our energies)
 so that the radial matrix elements of the same, we obtain a rough
estimate
for the ratio of matrix elements:
\beq
{\langle^3S_1 |{\cal O}_{\rm qm}| ^1P_1\rangle\over
\langle^3S_1 |{\cal O}_{\rm ic}| ^3P_1\rangle}  
\approx{2\sqrt{2}\over 3}
{ \delta m_N - \frac{\bar\delta m_N}{2} \over m_\pi}
\approx 0.03,
\eeq
using the quark model estimate
$\delta m_N - \bar\delta m_N/2 \simeq 4$ MeV. This comes from
taking $\delta m_N=3.1$ MeV\cite{jerry} so that
 $\bar{\delta}m_N=-1.8$ MeV.
This provides a first estimate of $A_{fb}$
in the percent range. (Cf. Eq. (\ref{deq}) for the triangle discrepancy.)

We turn now to the results of our detailed numerical evaluations. 
For illustration here we  again use
the quark model estimate
$\delta m_N -\bar\delta m_N/2 \simeq 4$ MeV. 
The 
 amplitudes arising from using this value in Eq.~(\ref{odeltam})
(denoted as loosely as ``qm''), 
as well as those obtained from the  sum of the previously
computed amplitudes of Ref.\cite{jouni} (denoted loosely as $\pi$$\eta$),
are listed in Table \ref{amplitudes}.
These are reduced matrix elements in the definition 
used {\it e.g.} in Ref. \cite{deShalit-Talmi}. 
Some overall kinematic factors are
ignored.

\begin{table}[tbh]
\caption{Real and imaginary parts
of amplitudes at $\eta=0.17$, $E=279.5$ MeV. The CSB amplitudes
have been multiplied by a factor of $10^{3}$. See the text for
the definition of qm and $\eta\pi$.}
\begin{center}
\begin{tabular}{|l|r|r|r|r|}
  \hline\hline
           & \multicolumn{2}{c|}{qm amplitude} 
           & \multicolumn{2}{c|}{$\pi$$\eta$ amplitude}
           \\ 
transition & real part & imag part  & real part & imag part\\ \hline
$^1P_1 \rightarrow s$ & 25.06 & $-$13.25&$-$10.38 & 6.34 \\
$^3S_1 \rightarrow p$ &  3.01 &   0.68& $-$3.86 &   $-$1.36 \\
$^3D_1 \rightarrow p$ &  0.82 &  $-$0.10& $-$7.26 &  2.65  \\
$^3D_2 \rightarrow p$ &  6.47 &   2.76 & 2.89 &   $-$0.3\\ 
$^3P_1 \rightarrow s$ &  0.901 &  -0.487& 0.901 &-0.487\\
$^1D_2 \rightarrow p$  & 0.535 & 0.086 & 0.535 & 0.086\\
\hline\hline
\end{tabular}
\end{center}
\label{amplitudes}
\end{table}

At 279.5 MeV, the  energy of the TRIUMF experiment,
the computed contribution from $\delta m_N$ and $\bar\delta m_N$
to the asymmetry of Eq.~(\ref{asym})
is found to be  0.97\%, or about -3.5 times  the strength  of 
(and opposite in sign to) the sum of the contributions 
calculated in Ref. \cite{jouni}. This earlier calculation
gave $A_{fb}$ =$-$0.28\%. As a consequence, the {\it total} asymmetry is
$A_{fb} = 0.69 \%$, three times the asymmetry 
of Ref. \cite{jouni}, with opposite sign.
This prediction is equivalent to a 10 standard deviation effect for
the experiment of Ref.~\cite{allena}.

While such a large result is anticipated from the estimates given above,
it is worthwhile to provide an   analysis of the amplitudes.
These results can be qualitatively understood by
looking at the interference of the most important
low partial wave amplitudes (denoted in the spectroscopic
notation by $l_{\pi J}$ or $l_{\pi LJ}$; the hat
indicates CSB amplitudes)
\begin{equation}
\sigma(\theta) - \sigma(\pi - \theta) =
4 \sqrt{\frac 2 3}\, Re\, p_2^* \hat s_1 \, \cos\theta
+ 2 \sqrt{2}\, Re\, s_{11}^* (\hat p_{22} - \sqrt{\frac 2 3}
\hat p_{01} + \sqrt{\frac 1 3} \hat p_{21} ) \cos\theta \, .
\label{ourcsb}\end{equation}
This can be compared with the charge-symmetric cross section
\begin{equation}
\sigma(\theta) + \sigma(\pi - \theta) =
2 |s_{11}|^2 + \frac 1 3 |p_2|^2 + |p_2|^2 \cos^2\theta\, .
\end{equation}
(The small amplitude $^1S_0 \rightarrow p_0$  can be
neglected in this qualitative comparison.)
For the case of qm, the largest two CSB amplitudes 
dominate so that one can clearly see that the
two interfering terms above add constructively.  The other 
CSB terms, computed earlier, 
have different phases and are therefore subject to
 destructive interference.

Note that the total cross section computed from the
present model at threshold yields an overestimate of the  data
by about 50\% 
\cite{comment}. We discuss how this could influence our present
result. 
The amplitude $s_{11}$ is
dominant, so one could arbitrarily reproduce the
data for the total cross section by dividing the amplitude
by $\sqrt{1.5}$. This amplitude enters in the 
second (smaller) term of 
Eq.~(\ref{ourcsb}) and in the denominator, 
so that the resulting value of
$A_{fb}$ (for all contributions) would increase by  a
factor of about  1.4. However, the same pion 
rescattering mechanism appears in the both  charge symmetric
and  CSB amplitudes. In that case, the error could cancel  out
in the ratio $A_{fb}$. We take an error estimate from 
the geometric mean
of unity and 1.4, which gives a theoretical uncertainty of
 about 20\%

As the isospin-violating amplitudes are small, they are approximately
linear in $\delta m_N -\bar\delta m_N/2$. Thus we are able to parametrize
$A_{fb}$ as a linear function of $\delta m_N -\bar\delta m_N/2$, at any energy.
At $E=279.5$ MeV, using the previous results 
we write  
\begin{equation}
A_{fb} \simeq -.28\%  \times \left(1-{0.87\over {\rm MeV}}
                     (\delta m_N -\frac{\bar\delta m_N}{2}) 
               \right)  .\label{afb}
\end{equation} 
The
predicted value of $A_{fb}$ depends strongly on the value of quantity
$\delta m_N -\bar\delta m_N/2$. 
Indeed, $A_{fb}$ nearly vanishes if
this is 1 MeV instead of 4 MeV. However,
there are estimates cited in the
reviews \cite{reviews,jerry,gl}
which would obtain 
$\delta m_N -\bar\delta m_N/2$ significantly greater  than 4 MeV.

The energy dependence of the amplitudes around this energy is
as expected ($s$ wave independent of momentum, $p$ wave
proportional to $q$). 
The asymmetry is not quite as good 
in its energy dependence as the elementary amplitudes:
below
300 MeV it is reasonably proportional to $q$; however, above
this energy it begins to curve down.
With the quark model estimate,
$A_{fb}$ increases to 1.76\% at 300 MeV and 1.87\% at 320 MeV. 

The principal conclusion of the present paper, based 
on Eq.~(\ref{afb}) is that 
the successful measurement of the value of the forward-backward asymmetry
of Eq.~(\ref{asym}) for the $np\to d\pi^0$ reaction will provide a serious test
of our understanding of QCD in terms of chiral effective field theory.
We look forward to the  results of Ref. \cite{allena} 
with great anticipation.

We close the paper by suggesting
that other experiments probing isospin violation
could supply additional information, in the whole providing
an even more stringent test.
We have already mentioned the need for more $\pi N$ data.
Another experiment \cite{andy},  planned at IUCF, 
would measure
the cross section for $dd\rightarrow \alpha \pi^0$.
Charge symmetry prevents the reaction $dd\to\alpha\pi^0$ from occurring, so
that
any non-zero cross section must be due to the breaking of the symmetry.
The cross section for  $dd\to\alpha\pi^0$ is proportional to the square of the
matrix element of the  CSB pion-production operator. Thus the cross section is
very small and hard to measure. However, there
is no interference with amplitudes
that respect charge symmetry, which would simplify the interpretation of
any measured cross section. Furthermore,
purely electromagnetic effects are very small, and 
 symmetries ---parity, angular momentum conservation and  
the identical nature of the two initial deuterons--- 
forbid the production of a
pion in a $p$ wave.  These features simplify the analysis of
 the reaction $dd\to\alpha\pi^0$.

\vspace{12pt}
\section*{Acknowledgments}
This research was supported in part by the Academy of Finland and
the Finnish Academy of Science and Letters through
the Vilho, Yrj\"o and Kalle V\"ais\"al\"a Foundation, 
the U.S. Department of Energy, and the U.S. National Science
Foundation. We thank C. Hanhart for useful comments. Two of
us (G.A.M and J.A.N.) thank ECT* where this work was completed.

\end{document}